# Stacked electrospun polymer nanofiber heterostructures with tailored stimulated emission

*Lech Sznitko,[a] Luigi Romano,[bc] Dominika Wawrzynczyk,[a] Konrad Cyprych,[a] Jaroslaw Mysliwiec,[a] and Dario Pisignano[*cd]*

[a] Faculty of Chemistry, Wroclaw University of Science and Technology, Wybrzeze Wyspianskiego 27, 50-370 Wroclaw, Poland

[b] Dipartimento di Matematica e Fisica "Ennio De Giorgi", Università del Salento, via Arnesano I-73100, Lecce, Italy

[c] NEST, Istituto Nanoscienze-CNR, Piazza S. Silvestro 12, I-56127 Pisa, Italy

[d] Dipartimento di Fisica, Università di Pisa, Largo B. Pontecorvo 3, I-56127 Pisa, Italy

[*] Corresponding author: dario.pisignano@unipi.it

ABSTRACT

We present stacked organic lasing heterostructures made by different species of light-emitting electrospun fibers, each able to provide optical gain in a specific spectral region. A hierarchical architecture is obtained by conformable layers of fibers with disordered two-dimensional organization and three-dimensional compositional heterogeneity. Lasing polymer fibers are superimposed in layers, showing asymmetric optical behavior from the two sides of the organic heterostructure, and tailored and bichromatic stimulated emission depending on the excitation direction. A marginal role of energy acceptor molecules in determining quenching of high-energy donor species is evidenced by luminescence decay time measurements. These findings show that non-woven stacks of light-emitting electrospun fibers doped with different dyes exhibit critically-suppressed Förster resonance energy transfer, limited at joints between different fiber species. This leads to obtain hybrid materials with mostly physically-separated acceptors and donors, thus largely preventing donor quenching and making much easier to achieve simultaneous lasing from multiple spectral bands. Coherent backscattering experiments are also performed on the system, suggesting the onset of random lasing features. These new organic lasing systems might find application in microfluidic devices where flexible and bidirectional excitation sources are needed, optical sensors, and nanophotonics.

# 1. Introduction

In the last decade, light-emitting polymer or organic/inorganic nanofibers have raised a continuously increasing interest, in view of their possible use in various fields including photonics.[1,2] Examples of recently demonstrated applications include the embedment of nanofibers in micro-lasers,[3-5] their use as turbid media for random lasing,[6,7] or as efficient materials for advanced lightening purposes.[8] Nanofibers realized by electrospinning[9-12] a method in which polymer solutions are ejected at very high strain rates by an external applied electric field, are especially suited to this aim. Advantages of electrospun fibers include the enhanced alignment of molecular backbones along the longitudinal axis of the organic filaments, which might lead to polarized emission from conjugated polymers,[13,14] the possibility of depositing coatings on large areas and with low cost, which makes the method compatible with manufacturing technologies for organic optoelectronics, and the high chemical versatility, which allows fibers to be realized with various polymer matrices, luminescent nanocrystals,[15] and light-emitting molecules. For instance, the variety of different organic dyes available for polymer doping makes the emission from electrospun fibers able to cover from the near ultraviolet to the near infrared without needing any complex chemistry or fabrication methods.[8,16,17]

However, while electrospun nanofibers can be arranged in various hierarchical architectures,[18-20] the configurations of light-emitting devices and lasers based on them, which have been tested so far, are still limited. Multilayer deposition by electrospinning has been performed by sequentially producing different populations of nanofibers onto the same collecting surface, which results in vertically-stacked heterostructures.[21] These multilayered fibrous meshes have been largely applied to obtain scaffold components for tissue engineering. In molecular electronics and photonics, stacking multiple organics in a networked structure might rise to enhanced charge transport, excitation migration, and multiple light scattering across the complex material. Due to the intimate and distributed interfaces between different polymer filaments, multicomponent electrospun heterostructures might find application within chemical reactors working at ultra-small scale ($10^{-21}$ mol),[22] and within light-emitting devices and lasers.

In this work, we demonstrate a hierarchical electrospun polymer architecture which exhibits light-emission and optical gain, composed by conformable layers with random two-dimensional organization of organic filaments, and vertical, three-dimensional compositional heterogeneity. The individual layers are doped with two different laser dyes, i.e. rhodamine 6G (Rh6G) and cresyl violet (CrV). These molecules could form a Förster resonance energy transfer (FRET) pair as acceptor and donor component upon polymer co-doping through blends. Such FRET effect was previously observed in blend microfibers.[23] Energy transfer mechanisms clearly cause stimulated emission (STE) to occur mainly from acceptors, and donor emission to be correspondingly quenched. These aspects lead to disadvantages in terms of spectral control, since multi-band STE in blend systems is hard to obtain and only associated to specific and narrow donor/acceptor relative concentrations. Similar conclusions can be drawn for other FRET pairs upon co-doping of polymer matrices, including thin-films and organic distributed feedback lasers.[24] Here, our approach relies on non-woven stacks of fibers with critically-suppressed FRET since acceptors and donors are mostly physically-separated, thus limiting donor-acceptor energy transfer events at joints between different fiber species, and making much easier to achieve simultaneous STE from multiple spectral bands.

The here presented systems show a significantly asymmetric optical behavior of its two sides, bichromatic STE which can be tuned in its spectral shape and threshold depending on the excitation direction, as well as incoherent random lasing features promoted by optical gain and interplaying light-scattering in the complex mat, as investigated by coherent backscattering (CBS) measurements. These organic light-emitting heterostructures might easily find application as novel excitation sources within lab-on-chips, sensors, and nanophotonic devices and architectures.

## 2. Experimental Section

Poly(methyl methacrylate) (PMMA, number-average molecular weight, $M_n$ = 120,000 g/mol), chloroform, and *N,N*-dimethylformamide (DMF) are purchased from Sigma-Aldrich. Rh6G and CrV are from Exciton. The dyes are separately dissolved in $CHCl_3$:DMF (4:1) solutions of PMMA (30% polymer/solvent w/v concentration). In both cases the concentration of dye is kept at 1% w/w with respect to PMMA, which is found to be adequate to provide good optical gain for reliable STE measurements. Electrospinning is carried out for a duration of 15 min and performed with a voltage bias of 18 kV applied over a distance of 25 cm from the 21 gauge needle of a syringe containing the solution to a copper target. Coverslips placed on the target are used for realizing samples suitable for the subsequent optical characterization. Applied flow rates during electrospinning are of 0.7 mL/h, achieved by a programmable syringe pump (Harvard Apparatus). Following the deposition of a first layer composed of Rh6G-doped filament, the second one (with CrV) is deposited with identical electrospinning parameters. The samples have areas of about 14 $cm^2$ and exhibit bright light emission under cw UV illumination, with different colors observed from the two sides composing the heterostructure (Fig. 1a-d).

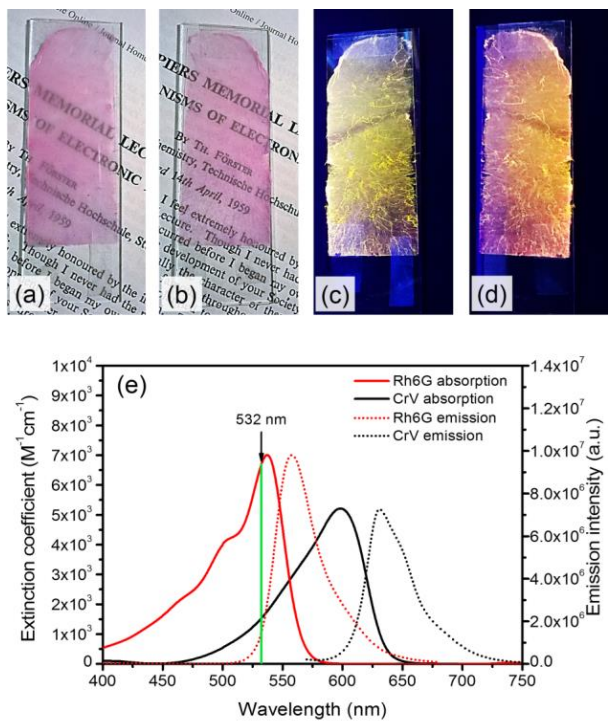

**Fig. 1.** (a-d) Photographs of the realized electrospun samples, under illumination with ambient light (a,b) and with UV light (c,d). The imaged side features fibers with either Rh6G (a,c) or CrV (b,d). (e) Absorption (left vertical scale, continuous lines) and fluorescence spectra (right scale, dotted lines) of Rh6G and CrV. THF solutions. The vertical, green line indicates the wavelength corresponding to the second harmonic of the Nd:YAG laser used for STE measurements in this work.

Further solutions of the two dyes are prepared in tetrahydrofuran (THF) for spectrophotometric and spectrofluorometric characterization of Rh6G and CrV (concentrations ≅ $10^{-4}$ M). Absorption spectra are obtained by a Jasco V-670 UV-VIS/NIR spectrophotometer. The absorption of Rh6G approximately ranges between 400 and 575 nm with maximum located at 537.5 nm, whereas the emission covers the yellow-orange part of the visible spectrum with maximum at 558 nm and Full Width at Half of Maximum (FWHM) equal to 40 nm. In CrV, the broad-band absorption spectrum approximately ranges between 450 and 650 nm with maximum at 598.5 nm, and the maximum of emission is located at 631 nm with FWHM equal to 42 nm (Fig. 1e).

Electrospun materials are inspected by an inverted optical microscope featuring a confocal system (A1R MP, Nikon), by a Hitachi F4500 spectrofluorometer, and by scanning electron microscopy (SEM, JSM 6610LVnx, JEOL Ltd). For confocal measurements, emission spectra are excited through a 20× objective (numerical aperture 0.50) with a cw $Ar^+$ laser operating at the wavelength ($\lambda$) of 514.5 nm. Micrographs are collected at different depths ($z_d$) into the sample (of thickness $d \cong 40$ μm), where $z_d \cong 0$ μm represents the position of the top (CrV) layer, whereas $z_d \cong 40$ μm refers to bottom (Rh6G).

CBS measurements are performed by the experimental set-up described in Ref. 25. STE from the realized heterostructures is excited by the second harmonic ($\lambda = 532$ nm) of a Nd:YAG laser (Surelite II, Continuum) operating at repetition rate of 10 Hz with pulse duration 5 ns, and analyzed by a spectral detection unit with a multi-anode photomultiplier. To control the intensity of pumping light, the excitation beam is driven through a half-wave plate and a Glan-Laser polarizer with vertical electric vector transmission. The beam of 3.3 mm diameter is incident perpendicular onto the plane of deposited fibers, and the emission is acquired from the sample using a USB 2000 fiber spectrometer (Ocean Optics, 4 nm resolution). Emission lifetime measurements are carried out by Time Correlated Single Photon Counting (TCSPC) technique, using a Becker & Hickl system that includes a TCSPC module (SPC-130-EM), a hybrid PMT detector (HPM-100-06) with detector control card (DCC 100) mounted on a Princeton Instruments spectrograph (Acton SpectraPro-2300i), and excitation delivered by a picosecond, 516 nm laser diode (BDL-516-SMC).

## 3. Results and Discussion

Fig. 2a shows confocal images of planes with different vertical coordinate ($z$) in the electrospun sample, as well as the correspondingly collected emission spectra. The yellow emission attributed to Rh6G dominates the fluorescence spectra from the firstly electrospun (bottom) layer (corresponding to $z_d \cong 40$ μm). When the focal plane is moved towards higher $z$ values (i.e., towards the top of the sample and at lower $z_d$ depth), the emission at the wavelength of 620 nm (attributed to the CrV fluorescence) emerges out of the overall spectra (Fig. 2b).

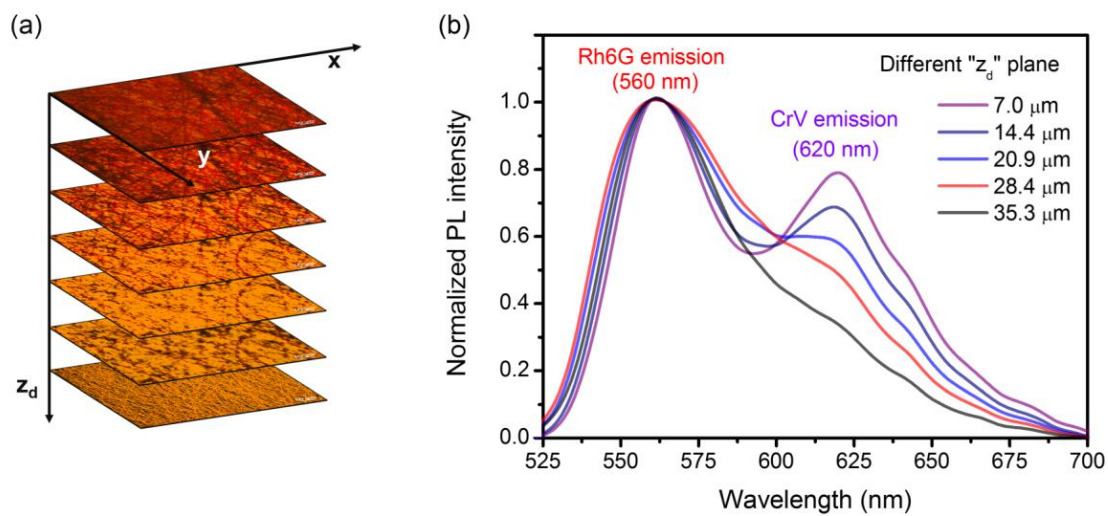

**Fig. 2.** (a) Confocal micrographs collected at different vertical coordinates $z_d$, representing the depth from top layer into the sample. (b) Corresponding emission spectra.

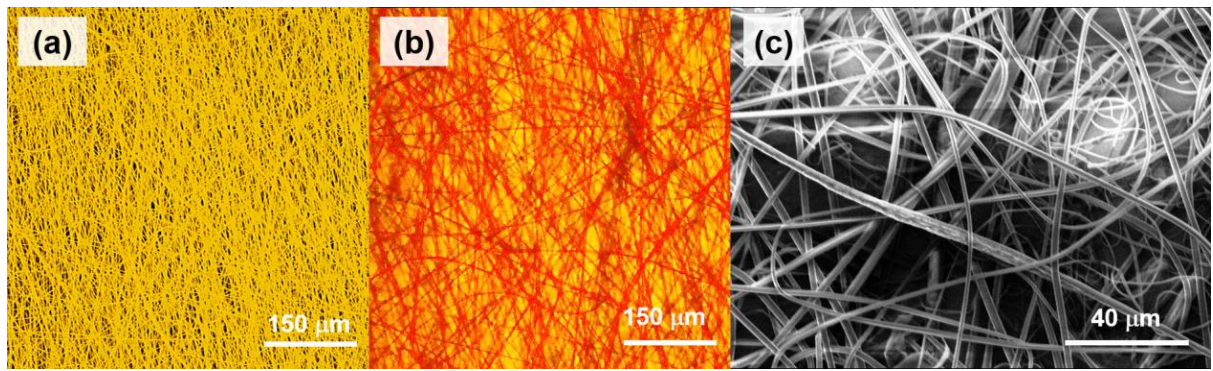

**Fig. 3.** Confocal micrographs of Rh6G-based and CrV-based fibers in the bottom (a) and top (b) layer of the hybrid electrospun heterostructures. (c) SEM micrograph of the typical fibrous morphology.

The top and bottom layers, imaged at higher magnification, are shown in Fig. 3a and 3b, respectively, which better highlight the different colors emitted from individual fibers composing the layers. The filaments mostly have ribbon shape with typical cross-section ($w \times h$) of 2.3-5.0 µm×1.1-2.0 µm, with no preferential orientation within each layer (Fig. 3c).

A marked asymmetric behavior is found for the light emission from the heterostructure. This is highlighted by excitation spectra of fibers which are measured at a given emission wavelength (660 nm, i.e. mainly occurring from CrV) while continuously varying the excitation wavelength (dotted lines in Fig. 4). When the system is excited from the CrV side, the band centered at the wavelength of 600 nm, referring to CrV absorption, is mostly accentuated in the excitation spectrum, while the Rh6G band is hidden and manifests as a shoulder on the short-wavelength tail of the spectrum. In the opposite situation, when excitation photons firstly impinge on Rh6G-based fibers, the excitation band consists instead of two maxima, namely a more intense attributed to Rh6G absorption (at 539 nm) and a second peak from CrV (again at about 600 nm).

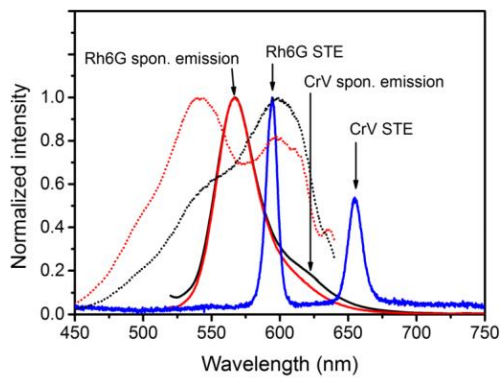

**Fig. 4.** Excitation and emission spectra of light-emitting electrospun heterostructures. Sample excitation is performed either on the Rh6G side (red lines) or on the CrV side (black lines). Solid lines represent fluorescence spectra excited by the wavelength of 480 nm, whereas dotted lines represent excitation spectra monitored at the wavelength of 660 nm. Blue curve: exemplary STE spectrum.

Similarly one would expect an influence of the direction of incidence of the excitation light on STE. Under sufficient pumping intensity from Nd:YAG second harmonic (532 nm), STE is found from the fibrous heterostructure at the wavelengths of 595 nm and 655 nm as displayed in Fig. 4 and in insets of Fig. 5a and 5b. The high-energy STE peak is associated with Rh6G whereas that at low-energy is due to CrV. More interestingly, when the excitation beam is firstly incident on the side of less absorbing fibers (i.e., with CrV dye), the fibers containing Rh6G molecules become efficiently pumped leading to a STE threshold fluence of (515±12) µJ/cm$^2$ (Fig. 5a). For sake of comparison, we recall that this value is a relatively low one compared to other systems with Rh6G described elsewhere.[25-28] This can be related to various concomitant and interplaying effects, such as moderate light-scattering of excitation photons through the fibrous structures[29] and photon waveguiding[30] within the complex mats, possibly gathering excitation priming STE on the laser dyes. The FWHM of emission is narrowed down to 10 nm above the pumping threshold as typically found for STE. The corresponding threshold fluence for the CrV STE band is of $(1061\pm6)$ $\mu J/cm^2$, and the FWHM for this band is reduced down to 13 nm.

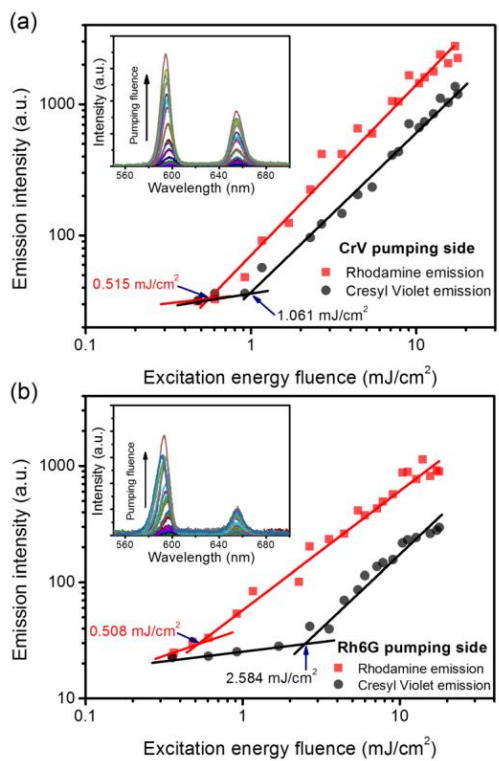

**Fig. 5.** Dependence of the emission intensity on the excitation energy fluence, for STE from Rh6G-based fibers (red lines) and from CrV-based fibers (black lines), respectively, and for excitation photons impinging on the heterostructure side made of CrV-based fibers (a) or on the side made by Rh6G-based fibers (b). Thresholds for STE are highlighted by arrows. Insets: emission spectra for varying pumping fluence.

When the sample is illuminated from the Rh6G side, the threshold for Rh6G STE does not change significantly $(508\pm9)$ $\mu J/cm^2$, whereas the threshold for the CrV STE more than doubles compared to the previous case, rising to (2584±5) µJ/cm$^2$ (Fig. 5b). The same value of Rh6G threshold found for different pumping sides clearly suggests that the extinction of directional excitation photons through CrV-based fibers does not affect Rh6G STE in a significant way. On the other hand, the significant increase of the threshold of CrV STE with pumping photons firstly impinging on Rh6G-based fibers indicates that Rh6G absorption is more effective in subtracting excitation photons before they can reach CrV-based fibers. Related to this, the relative intensity of the two STE peaks is also dependent on the pumping side. When fibers containing CrV dye are excited at first, the ratio of intensities between Rh6G and CrV bands is ~2.2, in the other case this value is reaches ~3.6. These findings agree with data shown in Fig. 1e, evidencing that the extinction coefficient for Rh6G at the wavelength of pumping laser (532 nm) is about four times larger than that of CrV. In both the excitation configurations, however, pumping of low-energy STE mediated by FRET from Rh6G-based fibers cannot be ruled out in principle, though limited at the interface between different species of fibers which involved junctions regions with a quite low density per heterostructure unit volume.

To analyze the STE mechanisms and the eventual effect of FRET more in depth, we measure involved emission lifetimes by the TCSPC technique. Emission decay curves are presented in Fig. 6a together with mono-exponential fits following deconvolution of Instrument Response Function (IRF). The efficiency of energy transfer process is determined through the equation:[31]

$$\eta = \left(1 - \frac{t_{DA}}{t_D}\right) \cdot 100\%, \tag{1}$$

where $t_{DA}$ is the fluorescence lifetime of donor (Rh6G) in presence of acceptor (CrV) and $t_D$ is the fluorescence lifetime without acceptor (only Rh6G). The obtained values of $t_{DA}$ are 2.46 or 2.41 ns for excitation photons impinging on CrV-based or on Rh6G-based fibers, respectively. A $t_D$ value equal to 2.67 ns is measured on a reference sample consisting of Rh6G-based fibers only. In this way, a FRET efficiency, $\eta = 10\%$, is found when excitation is carried out from the side with CrV-based fibers, and $\eta = 8\%$, for the excitation from the side with Rh6G-based fibers, confirming a marginal role of FRET in affecting the overall STE from the heterostructure. For this reason, it can be concluded that the STE from CrV when pumping photons reach the sample from the side made of Rh6G-based fibers (Figures 4 and 5a) is mainly due to residual light component at 532 nm or to reabsorption of photons emitted by rhodamine.

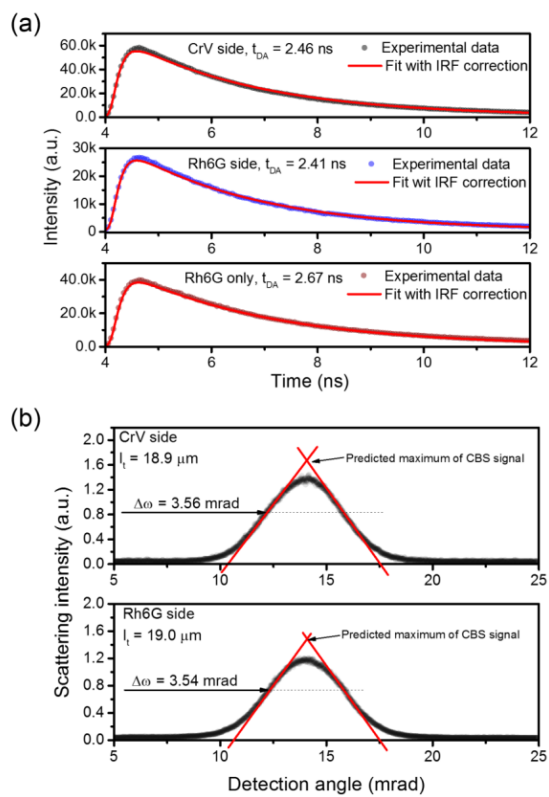

**Fig. 6.** (a) Spontaneous emission decay curves for the light-emitting electrospun heterostructure, when pumping is carried out from the side with CrV-based fibers (top panel) and with Rh6G-based fibers (middle panel), and for a reference sample containing only Rh6G-based fibers (bottom panel). (b) CBS signals measured for both the sides of the electrospun heterostructure. Δω: CBS FWHM.

Finally, we point out that the complexity of the bi-layered electrospun system suggests that the STE observed in our experiments is in fact incoherent random lasing.[32] To study this aspect, the transport mean free path ($L_t$) for photons in the material can be calculated by estimating the FWHM ($\Delta\omega$) of CBS angular dependencies as[33] $L_t = \frac{\lambda}{3\pi\Delta\omega}$, where $\lambda$ is the wavelength of scattered light. CBS measurements carried out for both sides of pumping lead to the value of 19 μm for the photon transport mean free path (Fig. 6b). The obtained $L_t$ well agrees with the hypothesis that emitted light would undergo multiple scattering when travels in the plane of fibers film, which would support incoherent random lasing in association with large excitation areas (spot diameter ≅ 3.3 mm). The excited region, being much larger than transport mean free path for photons, would naturally lead to a high number of overlapping optical modes.

The applications of asymmetric, light-emitting heterostructures made of electrospun fibers and showing multi-band STE and lasing could be numerous, including the development of low-cost sources laminated in transparent microfluidic devices featuring overlapping channels, of flexible, conformable or free-standing chemosensing surfaces with double-side optical response, and of tailored systems for parallel imaging. In addition, the compositional complexity of light-emitting electrospun heterostructures can be further increased, by depositing many layers featuring different fibers, to finely tune achieved colors and lasing threshold.

Published in RSC Advances: 8, 24175-24181. Doi: [10.1039/C8RA03640C](10.1039/C8RA03640C) (2018).## 4. Conclusion

We have realized a lasing heterostructure made of layered electrospun fibers. A textile material with complex light-emission and STE behavior is obtained, merging the optical features of two laser dyes embedded in different fiber species. The use of these fibers results in reduced quenching of a donor dye, therefore exciting the two species with the same wavelength leads to a donor emission not seriously affected by the presence of the acceptor, and to double-band STE from the system. The threshold of the emission as well as intensity ratio of STE bands are dependent on the pumping side in the layered geometry, showing that the spatial distribution of fibers also affects the optical properties of the obtained laser system. All of these experiments clearly indicates the significance of light-emitting fibers for realizing novel, low-cost media for random lasers and multi-band STE sources with versatile optical behavior. It can be concluded that a wide variety of dye-doped fibers and of electrospun heterostructures might be used for fabricating optical and laser materials, making possible to tailor light-scattering and optical gain properties in flexible molecular systems.

ACKNOWLEDGMENTS

This work was financially supported by The National Science Centre, Poland (2016/21/B/ST8/00468) and statutory funds of the Wroclaw University of Science and Technology. The research leading to these results has also received funding from the European Research Council under the European Union's Seventh Framework Programme (FP/2007-2013)/ERC Grant Agreements n. 306357 (ERC Starting Grant "NANO-JETS").